# Two-component model of a microtubule in a semi-discrete approximation

Slobodan Zdravković[a,c], Aleksandr N. Bugay[b], Slobodan Zeković[c], Dragana Ranković[d], Jovana Petrović[c,*]

[a] *Serbian Academy of Nonlinear Sciences, Beograd, Serbia*
[b] *Joint Institute for Nuclear Research, Joliot-Curie 6, 141980 Dubna, Russia*
[c] *Institut za nuklearne nauke Vinča, Univerzitet u Beogradu, 11001 Beograd, Serbia*
[d] *Farmaceutski fakultet, Univerzitet u Beogradu, 11221 Beograd, Serbia*

This article is dedicated to the memory of our colleague Slobodan Zeković, who passed away before publication.

[*] Corresponding author.
*E-mail address:* jovanap@vin.bg.ac.rs

*Keywords:*
Nonlinear dynamics of microtubules
Two-component model
Modulated solitary wave
Carrier and envelope velocities

ABSTRACT

In the present work, we study the nonlinear dynamics of a microtubule, an important part of the cytoskeleton. We use a two-component model of the relevant system. A crucial nonlinear differential equation is solved with semi-discrete approximation, yielding some localised modulated solitary waves called the breathers. A detailed estimation of the existing parameters is provided. The numerical investigation shows that the solutions are robust only if the carrier velocity of the breather wave is higher than its envelope velocity. That disproves the previously accepted solutions based on the equality of these velocities.

## 1. Introduction

A microtubule (MT) is a long, hollow cylindrical polymer spreading between a nucleus and cell membrane, representing a major part of the cytoskeleton [1]. Namely, intracellular protein filament networks exist in eukaryote cells. The two predominant types are filamentous actin (F-actin) and MTs [2, 3]. In this chapter, we deal with the latter only.

MTs play an essential role in the shaping and maintenance of cells and provide structural support to processes such as cell division [4]. They also serve as a network for motor proteins dynein and kinesin responsible for the transport of organelles and vesicles [5], which is probably the most interesting function from the standpoint of a physicist.

Here we assume that the structure of MT is known [2, 3]. We only point out that its surface is usually formed of 13 long structures called protofilaments (PFs), representing a series of electric dipoles called dimers. The MT mechanical behavior is ruled by the cylindrical shape of microtubules and the electrostatic interaction of the constituent dimers. The resulting nonlinear dynamics have been intensely investigated by intricate mathematical models over decades [6, 7, 8]. The models are mainly based on the fact that MT as a whole is ferroelectric [6, 9, 10] and typically assume one degree of freedom, either longitudinal or angular, depending on the coordinate used to describe the dimer`s oscillation, which has been conveniently used also in experiments [11].

Recently, we suggested a nonlinear two-component model [12]. This is an angular model. One angular coordinate determines the position around which the dimer oscillates, while the other one, also angular, describes the oscillation itself [12]. We assumed that the dimers oscillate in a tangential plane. Using a semi-discrete approximation we showed that a modulated solitary wave, known as a breather, moves along MT. Equality of the velocities of the envelope and carrier of the breather represents a so-called coherent mode (CM), a neat analytical solution that has been readily used during the past couple of years [13]. A crucial motivation for this article is to question CM. Using numerical calculations, we yield the important result showing that these velocities cannot be equal in a microtubule. We stress that the concept of CM presented here indicates the unlikeliness of the phase locking of the breather carrier and envelope and is different from the concept of a coherent energy transfer in tubulin studied in [14].

Notice that the modulated waves are present in many branches of physics. Examples include an electrical model of microtubules [15, 16], electrical engineering [17], plasma physics [18-20], solid state and condensed matter physics [21], geophysics [22], physics of fluids [23,24], quantum information [25], DNA biophysics [26-29], optics [30, 31], etc.

## 2. Two-component model and semi-discrete approximation

Interaction of a single dimer with surrounding ones that do not belong to the same PF can be modelled by W-potential energy [10, 12]

$$W(\theta) = -\frac{A}{2}\theta^2 + \frac{B}{4}\theta^4 - C\theta, \qquad (1)$$

where $\theta$ is the angle between MT axes and the direction around which the dimer oscillates. We assume that all the parameters in Eq. (1) are positive. The function $W(\theta)$ has two minima determining two stable positions for the dimer. These two positions correspond to the directions of electric fields $\vec{E}_1$ and $\vec{E}_2$, around which the dimer can oscillate. The resultant internal electric field $\vec{E} = \vec{E}_1 + \vec{E}_2$, coming from all dimers, is almost in the direction of PF. However, this is not stable and corresponds to the maximum of $W(\theta)$, while its "right" and "left" inclinations correspond to its minima [12].

We should not assume that the W-potential is reserved for MT analysis only. On the contrary, it is common in many branches of both classical and quantum physics [32-36].

Let us assume that all dimers are in ground states, which means that they belong to the deeper minimum of the W-potential. If we denote the appropriate field as $\vec{E}_1$ and the angle between it and PF as $\theta_0$, then the Hamiltonian for MT can be written as [12]

$$H = \sum_n \left[ \frac{I}{2}\dot{\varphi}_n^2 + \frac{k}{2}(\varphi_{n+1} - \varphi_n)^2 + W(\theta) - pE_1 \cos\varphi_n \right], \quad \theta = \theta_0 + \varphi, \tag{2}$$

where the coordinate $\varphi_n$ is the angle between the directions of the dimer at the position n and $\vec{E}_1$. Here, $I$ and $k$ are the moment of inertia of a single dimer and the inter-dimer stiffness parameter, respectively, while the dot means the first derivative with respect to time. Notice that the second term is the potential energy of the interaction between neighbouring dimers belonging to the same PF in the nearest neighbour approximation.

The last term in Eq. (2) comes from the fact that the dimer is an electric dipole existing in the field of all other dimers and $p$ is an electric dipole moment. It is assumed that $p > 0$ and $E_1 > 0$.

From Eqs. (1) and (2) and using Hamilton's equations of motion, we straightforwardly obtain the following dynamical equation of motion [12]:

$$I\ddot{\varphi}_n = k(\varphi_{n+1} + \varphi_{n-1} - 2\varphi_n) - A_0\varphi_n - C_0\varphi_n^2 - B_0\varphi_n^3, \tag{3}$$

where

$$A_0 = -A + 3B\theta_0^2 + pE_1, \quad B_0 = B - pE_1/6, \quad C_0 = 3B\theta_0. \tag{4}$$

We solve Eq. (3) using the semi-discrete approximation [37]. The method was explained in details in Refs. [26, 27]. Its mathematical basis is a multiple-scale method or a derivative-expansion method [38, 39].

According to the semi-discrete approximation, we assume small oscillations, i.e.,

$$\varphi_n = \varepsilon \Phi_n, \quad \varepsilon \ll 1, \tag{5}$$

and look for wave solutions of the form

$$\Phi_n(t) = F_1(\xi)e^{i\theta_n} + \varepsilon \left[ F_0(\xi) + F_2(\xi)e^{i2\theta_n} \right] + \text{cc} + O(\varepsilon^2), \tag{6}$$

$$\xi = (\varepsilon nl, \varepsilon t), \quad \theta_n = nql - \omega t. \tag{7}$$

Here, $\omega$ is the optical frequency of the linear approximation, $q = 2\pi/\lambda$ is the wave number, $l$ is the dimer's length, $n$ is an integer, cc represents complex conjugate terms, and the function $F_0$ is real. The point is that the function $F_1 \equiv F$ represents an envelope, which will be treated in a continuum limit, while $e^{i\theta_n}$, expressing discreteness, is a carrier component of the wave.

Following the procedure mentioned above [26, 27, 40], one can show that the function $F$ is a solution of the non-linear Schrödinger equation (NLSE)

$$iF_\tau + PF_{SS} + Q|F|^2 F = 0, \tag{8}$$

where the dispersion coefficient $P$ and the coefficient of nonlinearity $Q$ are

$$P = \frac{1}{2\omega}\left[\frac{kl^2}{I}\cos(ql) - V_g^2\right], \quad Q = \frac{1}{2\omega I}\left[2C_0(\mu - \delta) - 3B_0\right], \tag{9}$$

respectively, while the functions $F_0$ and $F_2$ can be obtained from $F$ as

$$F_0 = -\mu|F|^2, \quad \mu = 2C_0/A_0, \quad F_2 = \delta F^2, \quad \delta = C_0\left[4\omega^2 I - 4k\sin^2(ql) - A_0\right]^{-1}. \tag{10}$$

Here $S$ and $\tau$ are space and time coordinates [26, 27, 40], respectively, $V_g = d\omega/dq$ is the group velocity, and the appropriate dispersion relation is $I\omega^2 = 4k\sin^2(ql/2) + A_0$.

For $PQ > 0$, the solution of Eq. (8) is well known [26, 40-43], which finally gives the following expression for $\varphi_n(t)$:

$$\varphi_n(t) = 2A'\mathrm{sech}\left(\frac{nl - V_e t}{L}\right)\left\{\cos(\Theta nl - \Omega t) + A'\mathrm{sech}\left(\frac{nl - V_e t}{L}\right)\left[\frac{\mu}{2} + \delta\cos(2(\Theta nl - \Omega t))\right]\right\}, \tag{11}$$

where

$$A' = U_e\sqrt{\frac{1-2\eta}{2PQ}}, \quad L = \frac{2P}{U_e\sqrt{1-2\eta}}, \quad \Theta = q + \frac{U_e}{2P}, \quad \Omega = \omega + \frac{(V_g + \eta U_e)U_e}{2P}, \quad V_e = V_g + U_e. \tag{12}$$

The angle $\varphi_n(t)$ is shown in Fig. 3.

The solution of Eq. (8) comprises the parameters $U_c$ and $U_e$ that should satisfy $U_e > 2U_c$. It is convenient to use the parameters $U_e$ and $\eta$, defined as

$$\eta \equiv U_c/U_e, \quad 0 \leq \eta < 0.5. \tag{13}$$

## 3. Parameter selection

To roughly estimate moment of inertia of the dimer, we assume that it is an ellipsoid. Its approximate length and width are $8\,\mathrm{nm}$ and $4\,\mathrm{nm}$, respectively, [44] and, for the oscillations around the axes passing through the dimer's centre, we calculate $I = (m/5)(a^2 + b^2) = ml^2/16$, as $a = l/2$ and $b = l/4$.

It may be convenient to express the wavelength $\lambda$, defined through $q = 2\pi/\lambda$, as a multiple of the dimer's length $l = 8\text{nm}$, i.e., $ql = 2\pi/N$. This means that we should deal with either $q$ or $N$. In addition, we should estimate the values of the following physical parameters: $E_1$ (or $E_2$), $k$, $\eta$, $A$, $B$, and $C$. It is known that the electrical dipole moment strength is $p = 1.31 \times 10^{-27} \text{Cm}$ [12].

Let us first determine the parameters of the energy potential $A$, $B$, and $C$. Eq. (1) and the transformation $\theta = \sqrt{A/B}\,\psi$ yield

$$W(\psi) = \frac{A^2}{B}\left(-\frac{1}{2}\psi^2 + \frac{1}{4}\psi^4 - \sigma\psi\right) \equiv \frac{A^2}{B} f(\psi), \quad \sigma = \frac{C}{A\sqrt{A/B}}. \tag{14}$$

The function $f(\psi)$ appeared in Ref. [45] and its extreme values are

$$\psi_R = \frac{2}{\sqrt{3}}\cos F, \quad \psi_M = \frac{1}{\sqrt{3}}\left(-\cos F + \sqrt{3}\sin F\right), \quad \psi_L = -\frac{1}{\sqrt{3}}\left(\cos F + \sqrt{3}\sin F\right), \tag{15}$$

where

$$F = \frac{1}{3}\arccos\left(\frac{\sigma}{\sigma_0}\right), \quad \sigma_0 = \frac{2}{3\sqrt{3}}. \tag{16}$$

The letters R, L, and M denote the right and left minima, and the maximum.

The angles $\theta_{01}$ and $\theta_{02}$ were calculated in Ref. [46]. Hence, we can write

$$\sqrt{\frac{A}{B}}\psi_R = \theta_{01} = 1.0\,\text{rad}, \qquad -\sqrt{\frac{A}{B}}\psi_L = -\theta_{02} = 0.77\,\text{rad} \tag{17}$$

and easily obtain

$$F = 0.30\,\text{rad} = 17.3°, \quad \sigma = 0.62\sigma_0 = 0.24, \quad \psi_R = 1.10, \quad \psi_M = -0.26, \quad \psi_L = -0.85. \tag{18}$$

All this allows us to determine the values of the parameters $A$, $B$, and $C$. It is known that supply of energy from hydrolysis of guanosine triphosphate (GTP) in MTs may excite the vibrations [47]. This energy is $W_0 = 0.25\,\text{eV}$ and can be identified with the energy required for the dimer to escape from the well, which is, according to Eq. (14),

$$\frac{A^2}{B}\left[f(\psi_M) - f(\psi_R)\right] = W_0. \tag{19}$$

We picked $\psi_R$ because it corresponds to the deeper minimum, that is $f(\psi_R) < f(\psi_L)$, and straightforwardly obtain

$$A = 0.57 \text{eV}, \quad B = 0.69 \text{eV}, \quad C = 0.12 \text{eV}. \tag{20}$$

Let us estimate electric field strengths $E_1$ and $E_2$. As the resultant field is in the direction of PF, the x-components cancel out. Hence,

$$E_1 \sin\theta_{01} = E_2 \sin|\theta_{02}|; \quad E_1 \cos\theta_{01} + E_2 \cos|\theta_{02}| = E. \tag{21}$$

For $E = 1.7 \times 10^7$ N/C [48] we easily obtain $E_1 = 1.2 \times 10^7$ N/C and $E_2 = 1.5 \times 10^7$ N/C.

It was pointed out that we want to plot the angular displacement $\varphi_n(t)$. Let us concentrate on the choice of the minimum corresponding to $\psi_R$ and $E_1$. We can easily calculate $pE_1 = 1.6 \times 10^{-20}$ Nm $= 0.098$ eV as well as $A_{01} = 1.6$ eV, $B_{01} = 0.67$ eV, and $C_{01} = 2.1$ eV. Hence, we assign $A_0 \equiv A_{01}$, etc.

We now estimate the parameters $N$ and $\eta$. The product $PQ$ is an increasing function of $N$. In this article we pick $k = 12$ eV and, for this value, we can see that $PQ$ is positive for $N > 7.8$. The product $PQ$ is also positive in very narrow intervals for $N \leq 1$, but such wave would be extremely localized and these values are not acceptable. Notice that $\omega(N)$ has maximum, while both $P(N)$ and $Q(N)$ have minima for $N = 2$.

The parameter $\eta$ is probably the most interesting one. Let us define the amplitude $A_m$ and the wave width $\Lambda$ as $A_m = 2A'$ and $\Lambda = 2\pi L/l$, suggested by Eq. (11). Notice that $\Lambda$ is expressed in units of $l$. Fig. 1 shows how $A_m$ and $\Lambda$ depend on $\eta$ for $k = 12$ eV and $N = 14$. The function $A_m(N)$ has minimum at $N = 14$ for $k = 12$ eV and the assumed CM. The impact of $\eta$ on this result is negligible.

We see from Fig. 1 that the soliton amplitude decreases and the width increases with $\eta$. The assumption of small deviations from the mean microtubule position means that the solitons with large amplitudes must be ruled out. This excludes the cases of small $\eta$ with large $U_e$. However, even if we further exclude the extremely long solitons with thousands of dimers from consideration, the parameter region is still large enough to yield non-unique solutions with very different microtubule characteristics. For example, a CM wavepacket with the velocity $U_e = 134$ m/s and $\eta = 0.495$ has the same amplitude $A_m$ and length $\Lambda$ as an NCM with $U_e = 15$ m/s and $\eta = 0.1$, as shown in bold in Table 1. Therefore, additional analysis is needed to check the solution validity and identify a plausible parameter range.

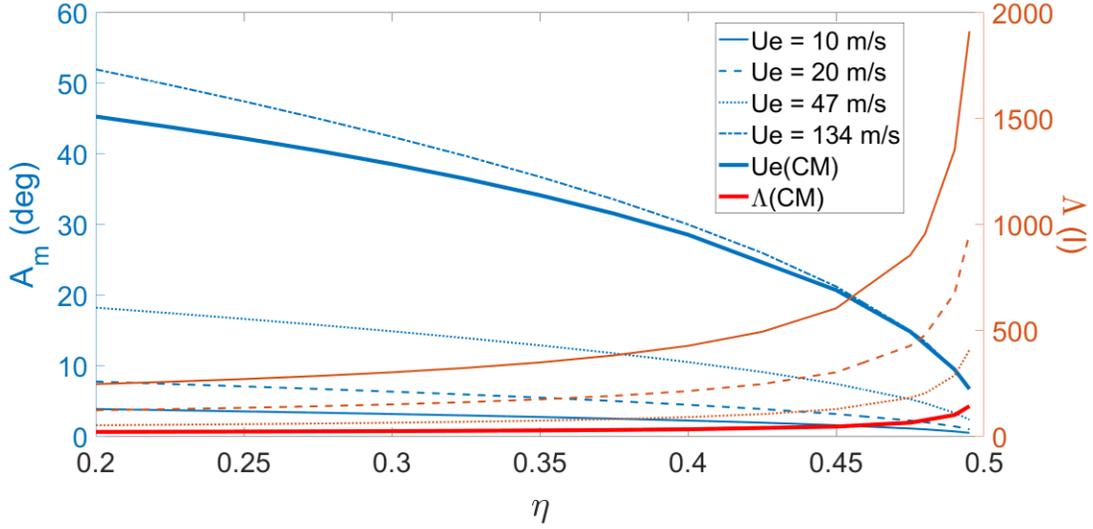

**Fig. 1.** The amplitude $A_m$ (left axis) and solitonic width $\Lambda$ (right axis) as functions of $\eta$ for $k = 12\text{eV}$, $N = 14$, and for different values of $U_e$ as given in legend. The solid lines show the CM case.

**Table 1.** The CM vs. NCM for $U_e = 15\,\text{m/s}$

| $\eta$ | $A_m$ (deg) | | $\Lambda(l)$ | | $U_e$ (m/s) | | $V_e$ (m/s) | |
|---|---|---|---|---|---|---|---|---|
| | CM | NCM | CM | NCM | CM | NCM | CM | NCM |
| 0 | 54 | 7.5 | 14 | 100 | 109 | 15 | 420 | 326 |
| 0.1 | 50 | **6.7** | 15 | **112** | 112 | 15 | 424 | 326 |
| 0.4 | 29 | 3.4 | 26 | 224 | 128 | 15 | 439 | 326 |
| 0.495 | **6.7** | 0.75 | **112** | 1001 | 134 | 15 | 445 | 326 |

## 4. Numerical investigations

We performed extensive numerical simulations to check the robustness of the obtained breathers for different MT parameters. Analytical solutions given by Eq. (11) were propagated numerically by solving the 2$^{nd}$ order ordinary differential Eq. (3) by Runge-Kuta RK4 method. In particular, we studied the dependence of the wavepacket length $L$ and envelope velocity $V_e$ on the parameters $\eta$ and $U_e$. We chose the widest parameter space, $0.1 < \eta < 0.5$ and $1\,\text{m/s} < U_e < 600\,\text{m/s}$, within the realistic MT modalities. Taking into account the estimated breather velocity of $445\,\text{m/s}$ (Table 1 for $\eta = 0.495$) and an average microtubule length of roughly 5µm [49], we concentrate on the time domain from 0 to 10ns. For completeness, we performed simulations up to 100ns propagation time and obtained the same conclusions. At the beginning of each simulation, readjustment of the wavefunction was observed, indicating that the breathers are not exact solutions, but that they faithfully describe dynamics in certain parameter regions.

Within the observed time interval, the form of the numerical solution reproduced well the breather form and analytically derived parameter dependences, for instance, the pulse width increased with $\eta$ and decreased with $U_e$. The pulse spread was extremely fast for $\eta \to 0.5$, where the breather width reached over thousands of dimers. On the other hand, for $U_e > 100\,\text{m/s}$ and $\eta < 0.45$ the wavepacket reduced to a sub-cycle pulse barely covering a pair of dimers.

In Fig. 2 we directly compare the breather velocities obtained numerically, $V_e^{\text{num}}$, and analytically by Eq. (11), $V_e^{\text{a}}$. $V_e^{\text{num}}$ was determined as the average velocity of the envelope maximum at the propagation time $t = 25\,\text{ns}$ (the range was extended to gain in accuracy). By applying this procedure to the analytical solution, the calculation error was estimated to be $< 1.5\,\text{m/s}$. We note that this error is purely numerical and is maintained across the parameters for which it could be reliably tested, i.e., for which the correct analytical solution existed.

The semidiscrete approximation (step $\Delta n = 1$ in RK4) rendered an excellent agreement within the numerical simulation for $U_e \leq 15\,\text{m/s}$, Fig. 3 (a). For faster breathers, numerical solutions consistently lagged behind their analytical counterparts; the leg increasing with the breather velocity (K dropping to 0.93 at $U_e = 50\,\text{m/s}$). Further, by allowing for smaller steps, $\Delta n \ll 1$, we found that $U_e = 50\,\text{m/s}$ is an upper limit of validity of the analytical breather solution in continous model, Fig. 3 (b)-(c). For even greater $U_e$, numerical pulse is significantly delayed and narrowed with respect to the analytical breather, indicating that the analytical solution does not capture the fastest dynamics.

We further illustrate the breather leg by plotting the dependence of the ratio of the numerically and analytically obtained velocities $K = V_e^{\text{num}}/V_e^{\text{a}}$ as a function of $U_e$ for constant η, Fig. 4a. The figure shows convergence of solutions for small $U_e$. Moreover, for fixed microtubule parameters reduction in the step $\Delta n$ results in convergence of $K$ to a constant value, for instance, to $K = 0.93$ for $U_e = 100\,\text{m/s}$ and to $K = 0.88$ for $U_e = 128\,\text{m/s}$. The latter is shown in Fig. 4b.

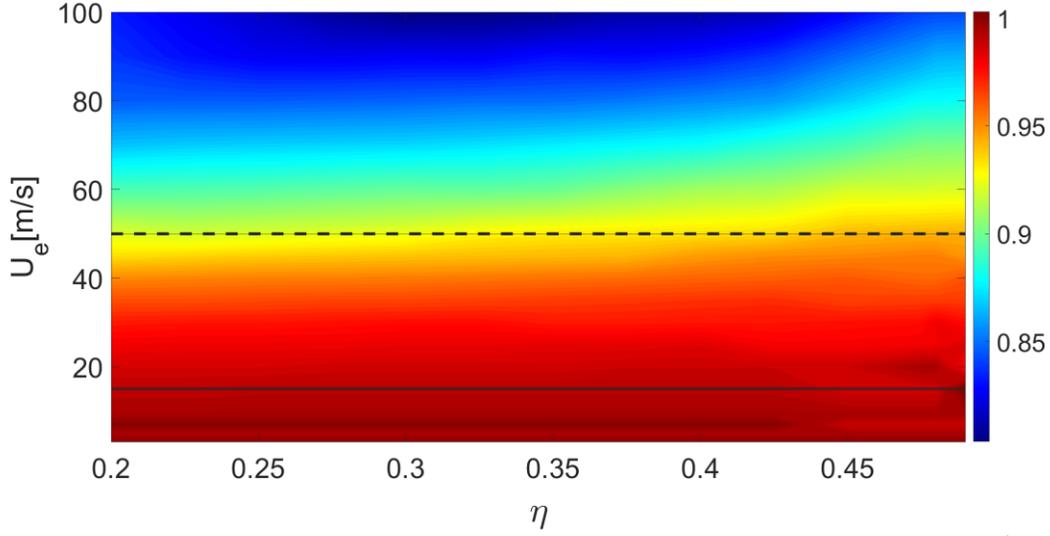

**Fig. 2.** The 2D colour map shows the ratio of the envelope velocities $K = V_e^{num}/V_e^a$ as a function of $U_e$ and $\eta$. Regions of validity of semi-discrete model (below the solid line) and continuous model (below the dashed line).

Finally, we revised the CM condition. Since the CM is a high-velocity breather ($U_e > 100\,\text{m/s}$ for all $\eta$), it is not validated by the above numerical envelope velocity criterion, Fig. 3(d). We corroborated this conclusion by finding that the carrier and envelope phases are not locked. However, we observe dependence in the form $V_e = r_{CE}\dfrac{\Omega}{\Theta}$, where $r_{CE}$ is the ratio between the envelope and carrier velocities ($r_{CE} = 1$ would indicate CM). We find that $r_{CE} < 1$ and that it only weakly depends on $\eta$ and $U_e$. For example, numerical estimates render $r_{CE} = 0.77 \pm 0.03$ in the range $40\,\text{m/s} < U_e < 60\,\text{m/s}$. This is an interesting result suggesting that the envelope velocity smaller than the carrier velocity may provide a robust breather.

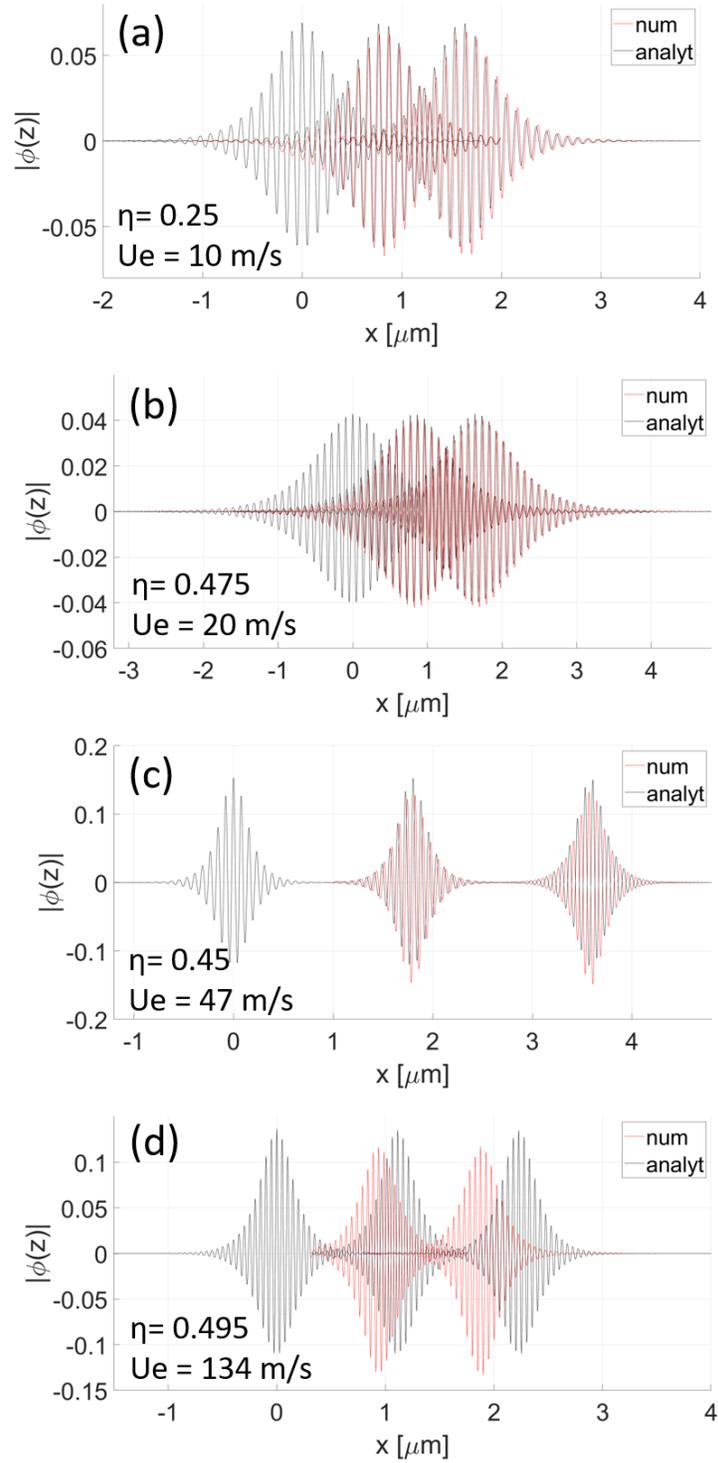

**Fig. 3.** The wavepackets obtained numerically (red line) and analytically (black line) for different values of parameters $U_e$ and $\eta$, given in plot legends. Pulses are shown at times 0, 0.5 ns, and 10 ns.

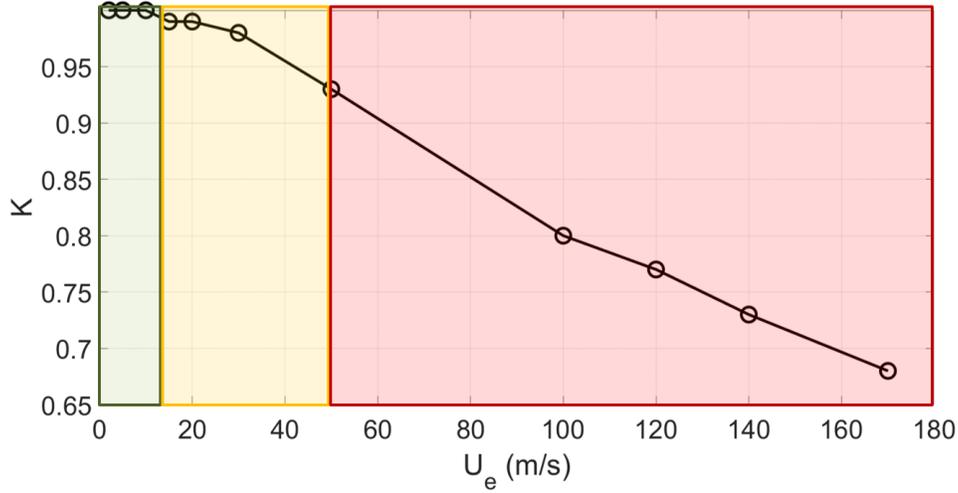

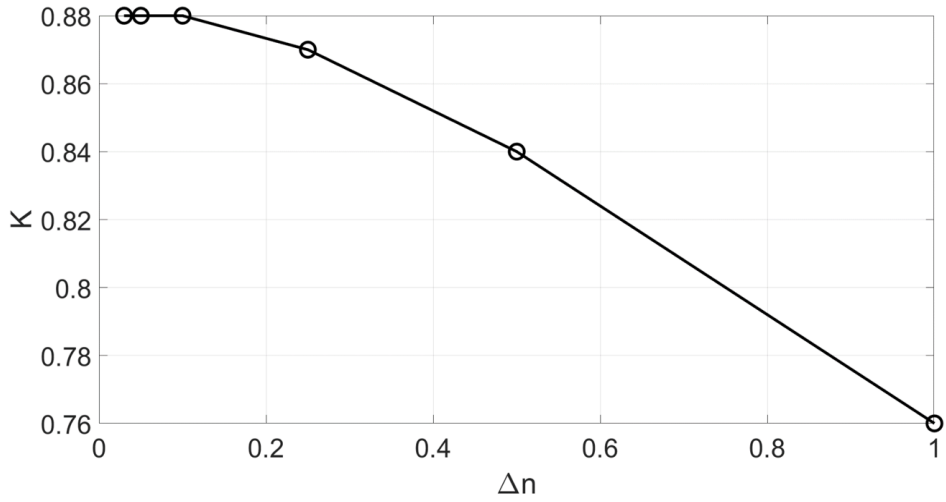

**Fig. 4.** Convergence of the ratio of the numerically and analytically obtained envelope velocity, K, with the velocity Ue evaluated in semi-discrete approximation (Δn=1). The green field shows the region of an excellent agreement between the semi-discrete and numerical solutions (K≥0.99), the yellow field the region of good agreement (K≥0.93), and the red field the region of poor agreement. In the yellow region, the analytical solution (Δn<<1) agrees well with the numerical solution. (b) Convergence of K with the numerical propagation step Δn evaluated for the assumed CM (Ue=127 m/s). In both plots η=0.4.

## 5. Conclusion

The nonlinear model explained here should be understood as a two component model in which one variable describes the oscillation of the dimer around the direction of the electric field and the other the orientation of the field. These possible orientations are determined by the minima of the electrostatic W-potential, which is the main source of nonlinearity in the model.

Solutions derived in semi-discrete approximation revealed that the dimers' oscillations follow breather dynamics along MT. Extensive numerical checks revealed that the validity of the semi-discrete, as well as the continuous analytical approximation, depends on the MT parameters. We performed a vast parameter analysis and derived realistic parameter ranges. However, parameter k is missing in the parameter selection. While the assumed value has physical sense and allowed us to plot the figures, the impact of this parameter on the MT dynamics and the validity of analytic and semi-discrete solutions will be studied in more detail as a part of the future work.

The main contribution of this work is discovery that only the solitons whose envelope velocity is lower than the carrier velocity have physical sense. The conclusion are relevant to the numerous MT models which render the analytically accessible soliton solutions, calling for attention when generalizing these solutions over the parameter space. The exact thresholds of the solution validity may depend on the MT model, i.e., the terms included in the NLSE. Finally, different approaches to discretization of the MT chain may be useful for explanations of MT dynamics, which is relevant to a broad range of physiological conditions and a general understanding of the human physiology [50].

**Acknowledgements**

D.R. and S.Z. would like to acknowledge the contribution of the COST Action CA21169, supported by COST (European Cooperation in Science and Technology). This work has been partially supported by the Project within the Cooperation Agreement between the JINR, Dubna, Russian Federation, and the Ministry of Science, Technological Development, and Innovation of the Republic of Serbia: Solitons and chaos in the nonlinear dynamics of biomolecules. J.P. and D.R. acknowledge support from the Ministry of Science, Technological Development and Innovation, Republic of Serbia, Grant numbers 451-03-66/2024-03/200017 and 451-03-65/2024-03/200161, respectively.